\documentclass[proof]{WileyASNA-v1}

\usepackage{graphicx}
\usepackage{wasysym}
\usepackage{array}
\usepackage{tikz}
\usepackage[latin1]{inputenc}

\articletype{Original article}%

\received{2 December 2020}
\revised{11 January 2021}
\accepted{14 January 2021}

\begin{document}

\title{Digital color codes of stars}

\author[1,2]{Jan-Vincent Harre}

\author[2,1]{Ren{\'e} Heller}

\authormark{Harre \& Heller}

\address[1]{\orgdiv{Institut f{\"u}r Astrophysik}, \orgname{Georg-August-Universit{\"a}t G{\"o}ttingen}, \orgaddress{\state{Friedrich-Hund-Platz 1, 37077 G{\"o}ttingen}, \country{Germany}}}

\address[2]{\orgname{Max-Planck-Institut f{\"u}r Sonnensystemforschung}, \orgaddress{\state{Justus-von-Liebig-Weg 3, 37077 G{\"o}ttingen}, \country{Germany}}}

\corres{Jan-Vincent Harre, \email{harre@mps.mpg.de}, Ren{\'e} Heller, \email{heller@mps.mpg.de}}

\abstract{Publications in astrophysics are nowadays mainly published and read in digitized formats. Astrophysical publications in both research and in popular outreach often use colorful representations of stars to indicate various stellar types, that is, different spectral types or effective temperatures. Computer generated and computer displayed imagery has become an integral part of stellar astrophysics communication. There is, however, no astrophysically motivated standard color palette for illustrative representations of stars and some stars are actually represented in misleading colors. We use pre-computed PHOENIX and TLUSTY stellar model spectra and convolve them with the three standard color matching functions for human color perception between 360\,nm and 830\,nm. The color matching functions represent the three sets of receptors in the eye that respond to red, green, and blue light. For a grid of main sequence stars with effective temperatures between 2300\,K and 55,000\,K of different metallicities we present the red-blue-green and hexadecimal color codes that can be used for digitized color representations of stars as if seen from space. 
We find significant deviations between the color codes of stars computed from stellar spectra and from a black body radiator of the same effective temperature. We illustrate the main sequence in the color wheel and demonstrate that there are no yellow, green, cyan, or purple stars. Red dwarf stars (spectral types M0V - M9V) actually look orange to the human eye. Old white dwarfs such as WD\,1856$+$534, host to a newly discovered transiting giant planet candidate, occur pale orange to the human eye, not white. Our freely available software can be used to generate color codes for any input spectrum such as those from planets, galaxies, quasars etc.}

\doi{10.1002/asna.202113868}

\jnlcitation{\cname{%
\author{J.-V. Harre} and
\author{R. Heller}} (\cyear{2020}), 
\ctitle{Digital color codes of stars}, \cjournal{Astron. Nachr.}, \cvol{2021:XXX-XXX}.}


\maketitle

\section{Introduction}
\label{sec:introduction}

Digitized representations of stars have become an integral part of publications in astronomy and astrophysics. Though realistic colors of stars are sometimes irrelevant for depictions of an astrophysical process, they can be helpful to support the message of a figure in a paper \citep{2010AsBio..10...89K,2010ApJ...720..904S,2013Natur.504..221K,2015ApJ...810...29H,2016AsBio..16..259H,2017AJ....154..230B} or a slide in an electronic science talk\footnote{\href{http://exoplanet-talks.org/}{http://exoplanet-talks.org}}. And in some cases, though the effect may be subtle, color is key, for example for the Rossiter-McLaughlin effect \citep{2012Natur.487..449S,2018Natur.553..477B} or for the color-induced displacement of binary stars \citep{2004A&A...423..755P}. In fact, the history of modern astronomy is deeply rooted in by-eye analyses of the apparent colors of stars and their relation to the temperature \citep{1911POPot..63.....H,1914PA.....22..275R,1925PhDT.........1P}. 


Beyond the scientific value of colorful stellar representations, the colors of stars play an important role in digitized science dissemination activities to the public, such as press releases\footnote{\href{https://www.eso.org/public/usa/news/eso1629/}{www.eso.org/public/usa/news/eso1629}}, educational websites\footnote{\href{http://www.exocast.org}{www.exocast.org}}, electronic books\footnote{\href{https://openstax.org/details/books/astronomy}{https://openstax.org/details/books/astronomy}} \citep{2019AAS...23422402F}, electronic slides in public talks etc.

That said, no uniform scale for the digitized color codes of stars exists. The digital illustration of stellar colors is often done in a pragmatic fashion rather than in a scientifically founded way and color representations of essentially the same types of stars differ greatly among publications in the various outlets listed above. Moreover, some myths about the red color of so-called red dwarf stars, of a brownish look of brown dwarfs, and of the white appeal of white dwarfs prevail. Here we develop a reproducible method to compute the colors of stars as they appear to the human eye (from space) and we report the digital color codes of main sequence stars.

\section{Methods}
\label{sec:methods}

\begin{figure*}[h!]
\centering
\includegraphics[width=0.75\linewidth]{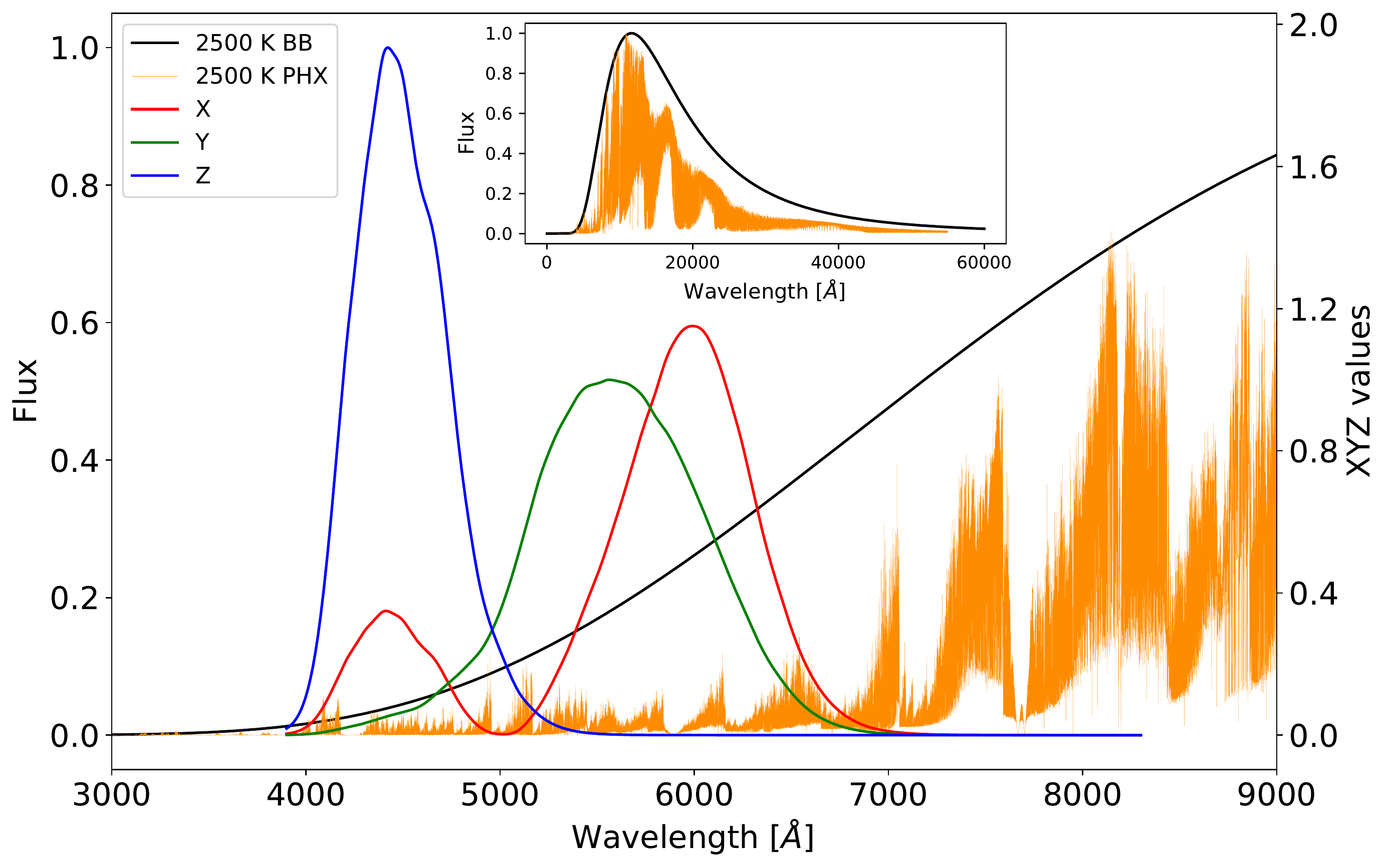}
\caption{Comparison of a black body spectrum (labeled BB, black solid line) and a PHOENIX spectrum of a main-sequence star (labeled PHX, orange line), both with an effective temperature of 2500\,K. The color matching functions are shown as X, Y and Z in red, green and blue, respectively. The inset shows an extended coverage of the spectra from 0 to $60,000$\,\AA.}
\label{fig:PHXvsBB}
\end{figure*}

\subsection{Stellar model spectra}

As input spectra to our color calculations, we used pre-computed model stellar spectra of the flux densities as a function of wavelength. For effective temperatures $2300\,{\rm K}\,\leq\,T_{\mathrm{eff}}\leq12,000$\,K we used spectra from the publicly available\footnote{\href{http://phoenix.astro.physik.uni-goettingen.de/}{http://phoenix.astro.physik.uni-goettingen.de}} PHOENIX spectral library of \citet{2013A&A...553A...6H}. Additional PHOENIX spectra with $T_{\rm eff}$ up to 15,000\,K were provided by courtesy of T.~O.~Husser in private communication. The PHOENIX model grid embraced surface gravity values $0\leq\log(g)\leq6$ ($g$ expressed in units of cm\,s$^{-2}$) and stellar metallicities $[{\rm Fe/H}]\in\{0,-1,-2\}$. All spectra were provided on the same wavelength grid (500\,\AA$\,\leq\lambda\leq5.5\,\mu$m) and with a typical resolution of $R=\lambda/{\Delta}\lambda=500,000$ in the optical regime.

These spectra were computed using version 16 of the PHOENIX software \citep{1999JCoAM.109...41H}. Spherical symmetry of the stars was assumed and each atmosphere was represented by 64 layers. The assumption of local thermal equilibrium (LTE) was justified by the maximum effective temperatures of 12,000\,K, and dust settling was ignored because it is insignificant for $T_{\rm eff}\geq2300$\,K. For details we refer the reader to \citet{2013A&A...553A...6H}.

For OB stars with $16,000\,{\rm K}\,\leq\,T_{\mathrm{eff}}\leq55,000$\,K we used TLUSTY models\footnote{\href{http://tlusty.oca.eu/Tlusty2002/tlusty-frames-OS02.html}{http://tlusty.oca.eu/Tlusty2002/tlusty-frames-OS02.html}} of the BSTAR2006 grid of \citet{2007ApJS..169...83L} and the OSTAR2002 grid of \citet{2003ApJS..146..417L}. These models take into account the effects of non-LTE in plane-parallel, hydrostatic atmospheric layers. Due to the very weak absorption lines of these hot stars, the differences in the color codes are very small across different stellar metallicities. As a consequence, we restricted our analysis to solar metallicity.

\subsection{Color codes and Matching functions}

To infer the digital color codes for the model spectra as if they were perceived by the human eye, we used {\tt color\_system.py}\footnote{\href{https://scipython.com/blog/converting-a-spectrum-to-a-colour/}{https://scipython.com/blog/converting-a-spectrum-to-a-colour}}. This python module uses data from color matching functions (CMFs), which describe the perception of light by the cone cells in the human eye under standardized illumination, to convert a spectrum to a color code. Since the model stellar spectra have a high wavelength resolution we used the latest values of the 2-degree XYZ CMFs\footnote{\href{http://cvrl.ucl.ac.uk/}{http://cvrl.ucl.ac.uk}}, transformed from the 2006 CIE (International Commission on Illumination) 2-degree LMS (long-medium-short color space) cone fundamentals \citep{Stockman2008PHYSIOLOGICALLYBASEDCM}, which come with one data point per {\AA}. We then degraded the resolution of the synthetic stellar spectra to the same wavelength resolution.

In Fig.~\ref{fig:PHXvsBB} we show the CMFs (blue, green, red lines) together with two example spectra, one being a black body of 2500\,K (black line) and one being a PHOENIX spectrum for $T_{\rm eff}=2500$\,K, $\log(g)=5$, and $[{\rm Fe/H}]=0$. Both spectra are normalized to a value of one (see inset). The presence of strong molecular absorption bands in the PHOENIX spectrum suggests that the resulting color perception, and therefore also the digital color codes, differ substantially between a black body and a stellar spectrum of the same effective temperature.




\subsection{Spectral typing}
\label{sec:SpT}

The PHOENIX spectra are not formally linked to a given stellar spectral type (SpT), although empirical relationships can be used to relate PHOENIX spectra with SpT. To simplify the choice of an appropriate color representation for a given SpT, one of our aims was to create a look-up table that is astrophysically motivated but particularly suitable from an astronomer's perspective.

To this end, we matched spectral types of main-sequence stars with the corresponding $T_{\rm eff}$ and $\log(g)$ values from our grid of PHOENIX models -- and for consistency also for the TLUSTY models. First, we used Table~5 of \citet{2013ApJS..208....9P} to match $T_{\rm eff}$ from the synthetic models with observed spectral types of stars. These authors used a weighting scheme of a large sample of standard main-sequence stars from the literature to infer a $T_{\rm eff}$-SpT relation for $T_{\rm eff}~{\leq}~34,000$\,K. Stellar surface gravities were not provided. As an alternative, we define three intervals of $\log(g)$ based on the $T_{\rm eff}$-$\log(g)$ relation derived from a sample of detached eclipsing stellar binaries \citep{2018MNRAS.479.5491E}:

\begin{align}  \nonumber
        & ~T_{\rm eff}~\leq~3648\,{\rm K} & \Rightarrow \ \ \ \ \ \ \ \ \ \ \log(g) = 5.0\\ \nonumber
3648\,{\rm K}~< & ~T_{\rm eff}~\leq~6152\,{\rm K} & \Rightarrow \ \ \ \ \ \ \ \ \ \ \log(g) = 4.5\\
6152\,{\rm K}~< & ~T_{\rm eff}            & \Rightarrow \ \ \ \ \ \ \ \ \ \ \log(g) = 4.0
\end{align}

For $T_{\rm eff}>34,000$\,K we used Table~7 of \citet{2018MNRAS.479.5491E} for the SpT-to-$T_{\rm eff}$ matching.

\subsection{Limb darkening}

For illustration purpose only, we generate a library of plots of the apparent stellar surface that include limb darkening as parameterized by the quadratic limb darkening law,

\begin{equation}\label{eq:LD}
\frac{I(\mu)}{I(1)} = 1 - a(1-\mu) - b(1-\mu)^2 \ ,
\end{equation}

\noindent
where $\mu~=~\cos(\gamma)$, $\gamma$ is the angle between the line of sight and the normal to the stellar surface, $I$ is the specific intensity, and $(a,b)$ are the limb darkening coefficients. This model does not take into account the possible wavelength dependence of limb darkening, that is, chromatic effects. We used limb darkening coefficients that reproduce the limb darkening as observed in the $G$ filter of the Gaia mission \citep[Table 2 in][]{2019RNAAS...3...17C}. We chose the Gaia $G$ filter because its response curve matches reasonably well with the response function of the human eye \citep{2018A&A...617A.138W}. Given that the PHOENIX and TLUSTY model grids and the limb darkening tables have different step sizes in the model parameters, we used linear interpolation of the limb darkening coefficients whenever necessary to represent stars on the model grid.


In technical terms, limb darkening was implemented into our computer code by first creating a two-dimensional array with $1001~{\times}~1001$ entries, each containing a sub-array with the respective linear RGB color triple values of the star. Each linear RGB triple entry corresponded to a radial distance from the stellar center. Using the relation $\mu=\sqrt{1-r^2}$ \citep{2019A&A...623A.137H}, with $0~{\leq}~r~{\leq}~1$ being the radial distance from the star disk in units of stellar radii, each entry in the color array was multiplied with the corresponding intensity of the limb darkening in Eq.~\eqref{eq:LD} law to obtain the radial limb darkening profile.

\begin{figure*}[t]
\centering
\includegraphics[width=0.4\linewidth]{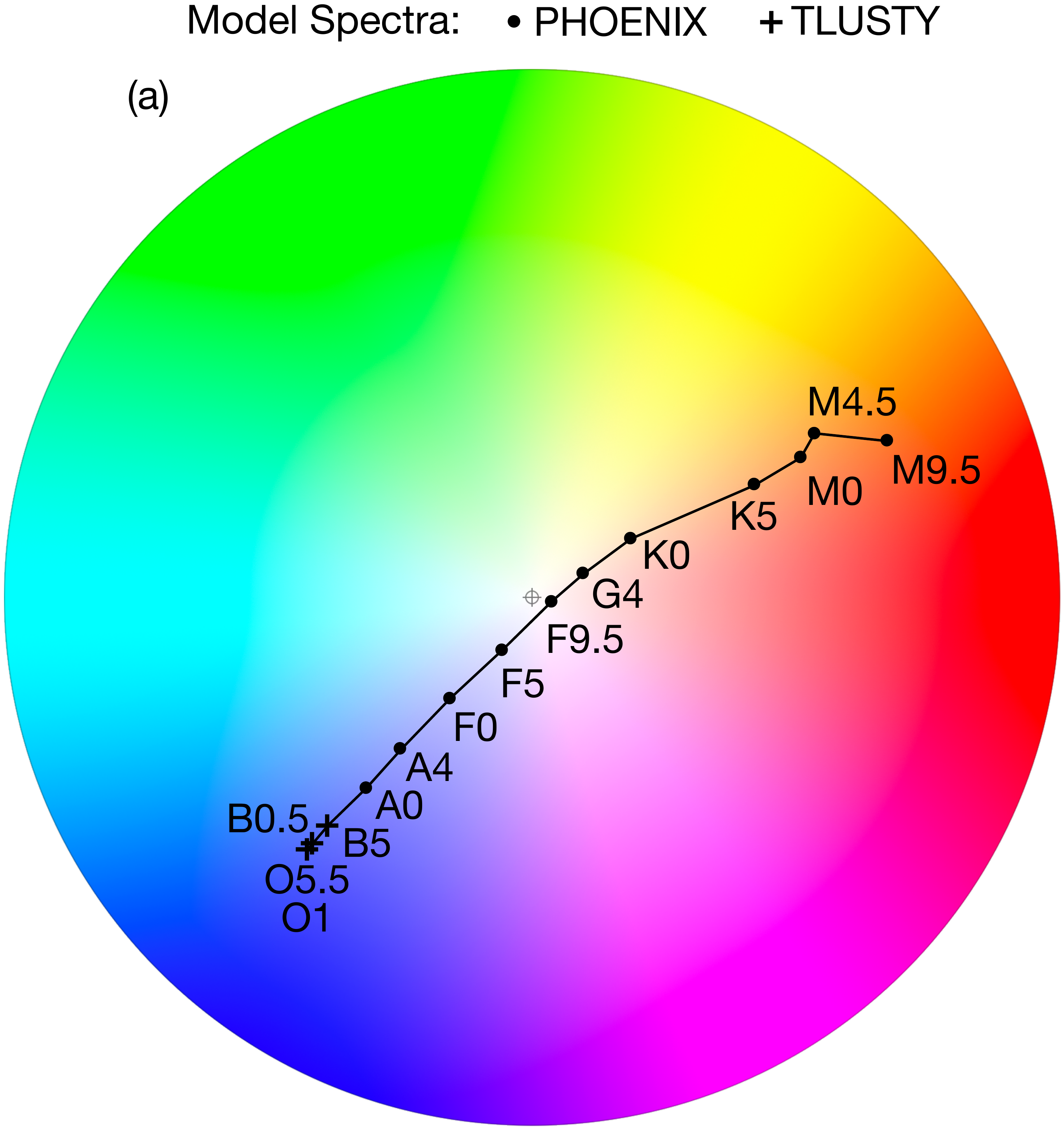}
\hspace{0.4cm}
\includegraphics[width=0.4\linewidth]{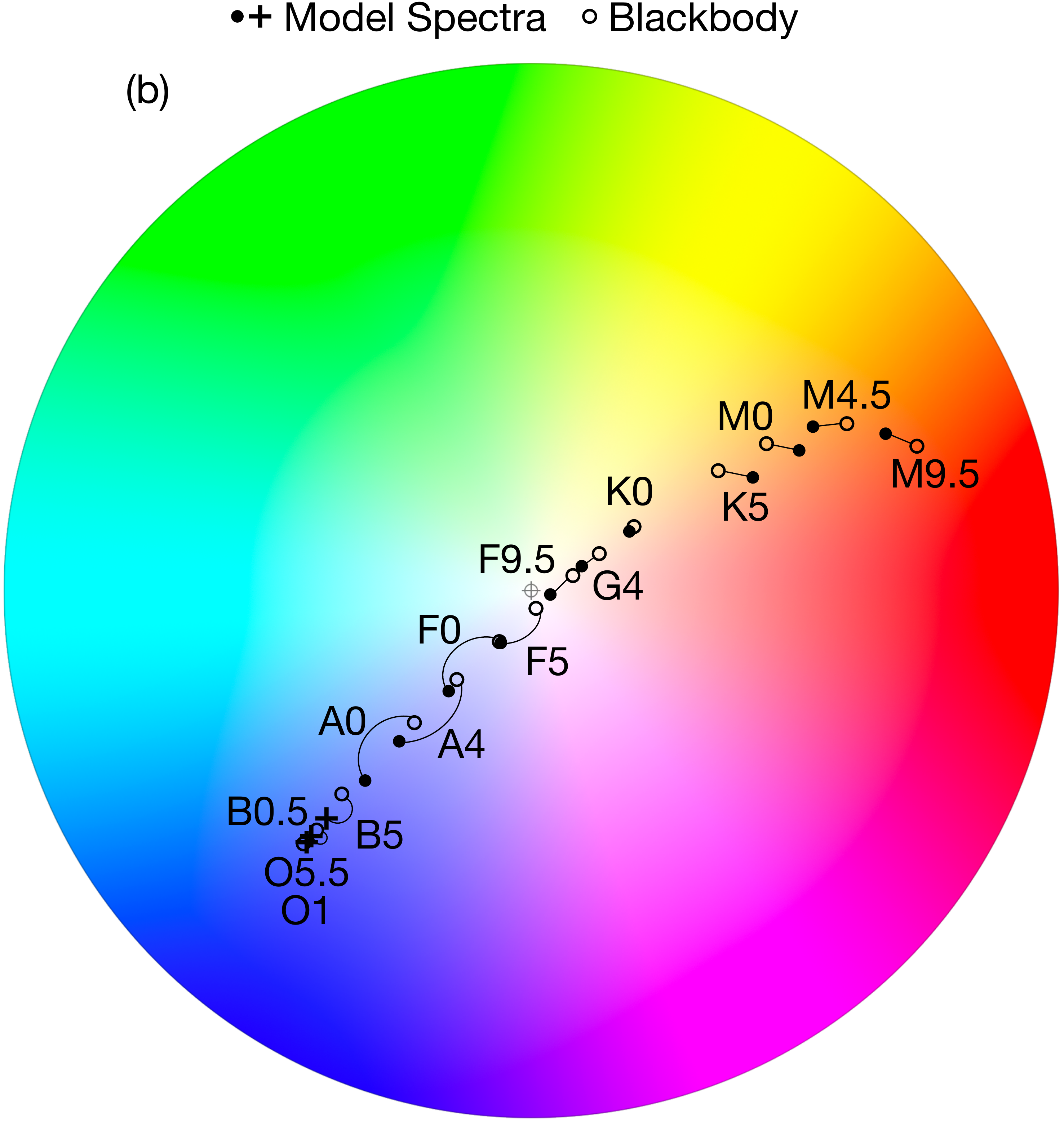}
\caption{(a) Main sequence stars in the color wheel. Colors are derived from synthetic stellar spectra. (b) Comparison of colors derived from synthetic spectra (filled circles) with colors calculated from black bodies of the corresponding effective temperatures (open circles). Spectral types are indicated along the curves.}
\label{fig:wheel}
\end{figure*}

\section{Results}
\label{sec:results}

\subsection{Digital color codes}

In Table~\ref{tab:colors_bb} we list the linear RGB and Hexadecimal (Hex) color codes of black bodies with $2300~{\leq}~T~{\leq}~12,000$\,K, the temperature range of which corresponds to the publicly available PHOENIX models. Columns 2 and 3 refer to the colors of a black body in vacuum without any extinction or transmission through an additional medium between the radiator and the observer.

In Tables~\ref{tab:colors_Z0} - \ref{tab:colors_Z-2} we show the linear RGB and Hex color codes as computed from the PHOENIX spectra for $2300~{\leq}~T_{\rm eff}~{\leq}~12,000$\,K. The color variation of stars with $T_{\rm eff}>12,000$\,K is hardly notable to the human eye, which is why we restricted Tables~\ref{tab:colors_Z0} - \ref{tab:colors_Z-2} to $T_{\rm eff}~{\leq}~12,000$\,K. Table~\ref{tab:colors_Z0}, for $[{\rm Fe/H}]=0$, contains color codes for 841 PHOENIX spectra; Table~\ref{tab:colors_Z-1} and its 846 entries refer to $[{\rm Fe/H}]=-1$; and Table~\ref{tab:colors_Z-2} is a list of 848 PHOENIX spectra assuming $[{\rm Fe/H}]=-2$. Table~\ref{tab:SpT_colors} is a look-up table for the color codes of main-sequence stars of a given spectral type and embraces both the PHOENIX and the TLUSTY models with effective temperatures up to 55,000\,K.

Figure~\ref{fig:wheel} is a visual representation of these color codes. Panel (a) shows the stellar main sequence in the color wheel. Color codes were first computed using the PHOENIX and TLUSTY spectra and then a selection of spectral types were marked in the color wheel in steps of about half a spectral class between M9.5 to O1 (all of luminosity class V). To our knowledge, this is the first digital color code representation of the long known astronomical observation that there are no yellow, green, cyan, or purple stars. At the lowest temperatures, the main sequence begins with orange stars of SpT M9.5, then continues to produce brighter and brighter stars across the K class until it reaches the white point with stars of SpT F9.5. Sun-like stars with SpT G2 are slightly yellowish. Early-type stars of spectral classes A, B, and O tend to have ever more blue colors. Interestingly, for spectral types earlier than about B5 the colors of increasingly hotter stars converge towards a blueish tone of linear RGB = (90,123,255).

In Figure~\ref{fig:wheel}(b) we illustrate the differences between the colors of main-sequence stars as depicted in panel (a) and black bodies of the same effective temperatures for the same selection of spectral types. Pairs of model spectra (filled circles and crosses) and black bodies (open circles) of the same $T_{\rm eff}$ are connected with a black line. Interestingly, while there are significant deviations between the two models for both the most late-type stars and the most early-type stars, colors virtually agree for SpT K0. We also note two inflection points in the color differences, one between M4.5 and M0 and a second one between K5 and K0.

\begin{figure}[t]
    \centering
    \includegraphics[width=0.38\linewidth]{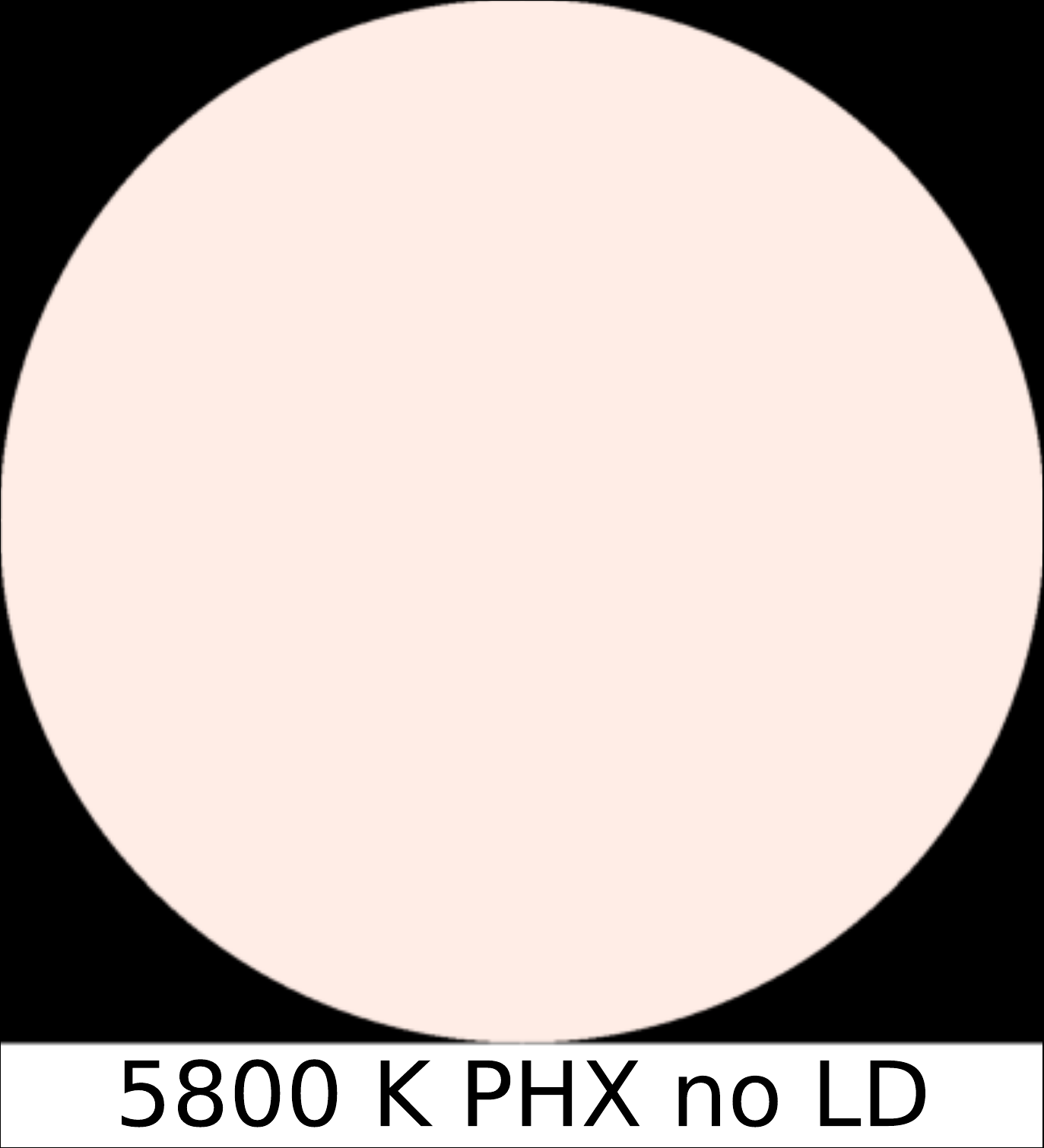}
    \hspace{1.3cm}
    \includegraphics[width=0.38\linewidth]{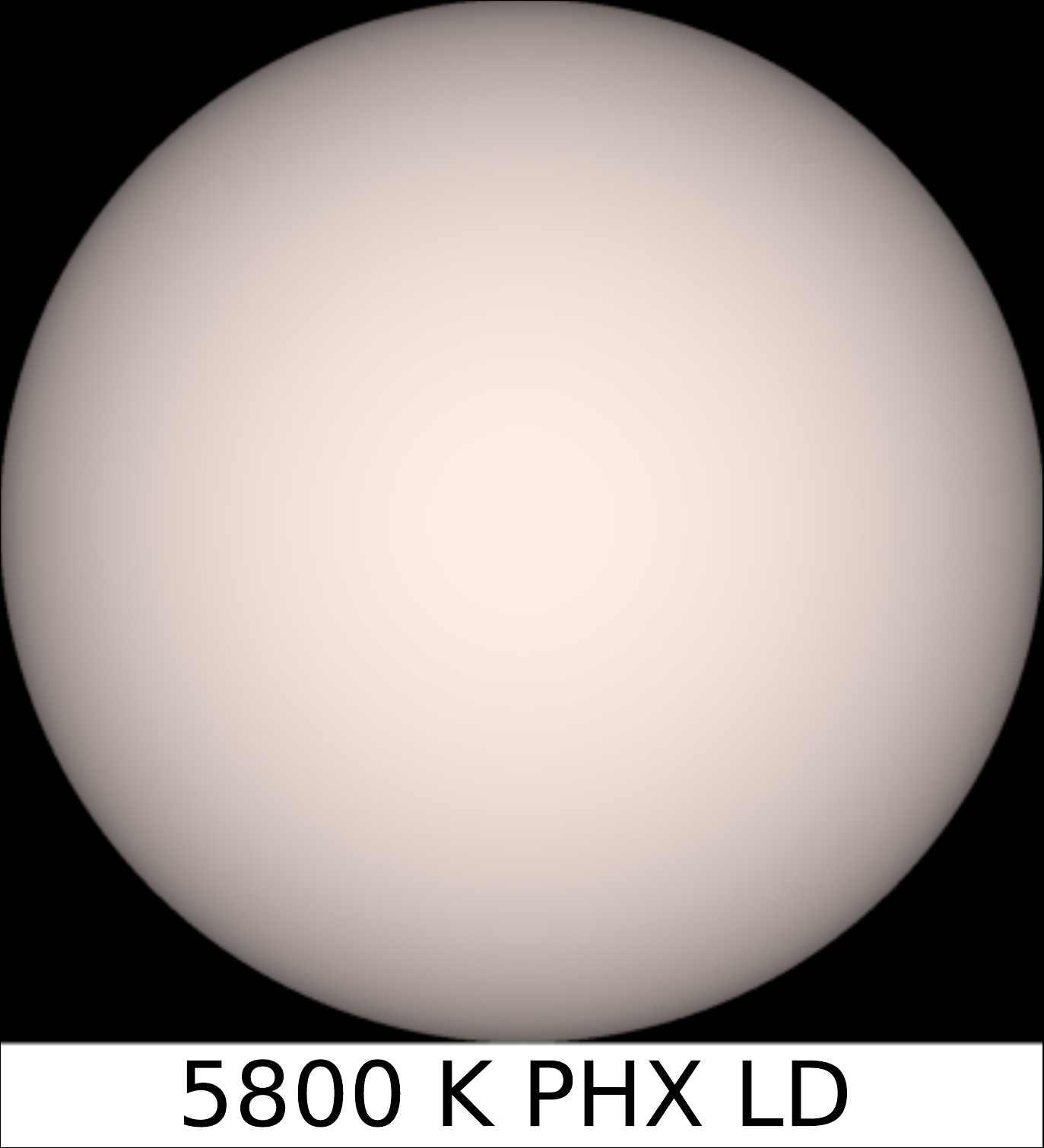}
    \caption{Comparison of the same star with and without limb darkening. For the computation of the color we chose the PHOENIX spectrum of a sun-like star with $T_\mathrm{eff}~=~5800\,$~K, $\log(g)~=~4.5$, and $[{\rm Fe/H}]~=~0$.}
    \label{fig:ld_comparison}
\end{figure}

\subsection{Limb darkening}

The effect of limb darkening is illustrated in Fig.~\ref{fig:ld_comparison} for a sun-like star. The left panel has no limb darkening, the right panel includes the quadratic limb darkening law. Our implementation of limb darkening comes after the calculation of the linear RGB codes from the model spectra, which means that the linear RGB colors codes given in Tables~\ref{tab:colors_Z0} - \ref{tab:colors_Z-2} refer to the center of the stellar disk.

For each of the 107 stars listed in the look-up Table~\ref{tab:SpT_colors}, we generated an illustration in portable document format (PDF, version 1.4) using the python plotting package {\tt matplotlib} that is similar to the one in the right panel of Fig.~\ref{fig:ld_comparison}. These electronic images are available in the online version of this report.

\subsection{Black bodies and stellar spectra}

\begin{figure}
\centering
\includegraphics[width=0.38\linewidth]{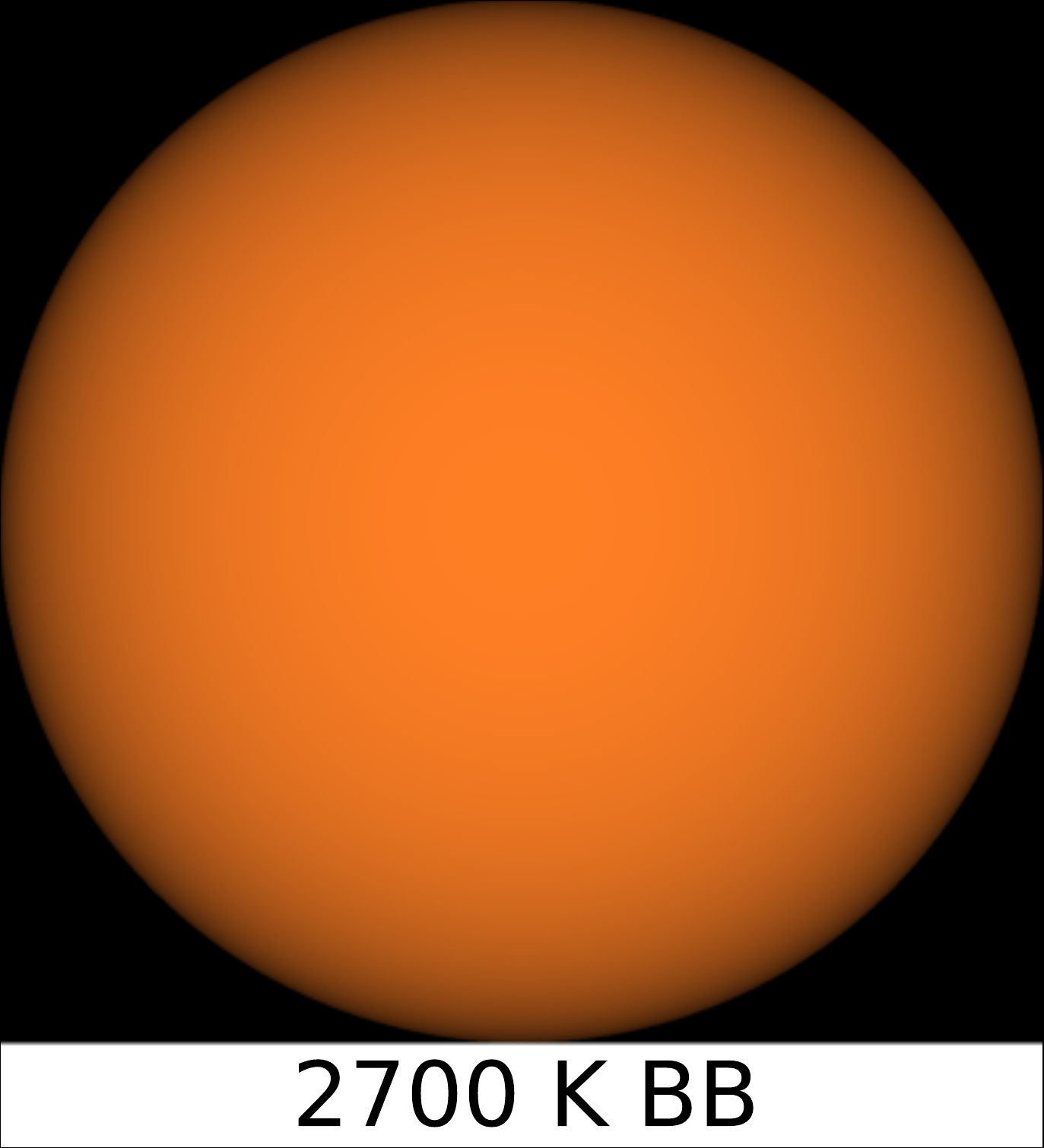}
\includegraphics[width=0.38\linewidth]{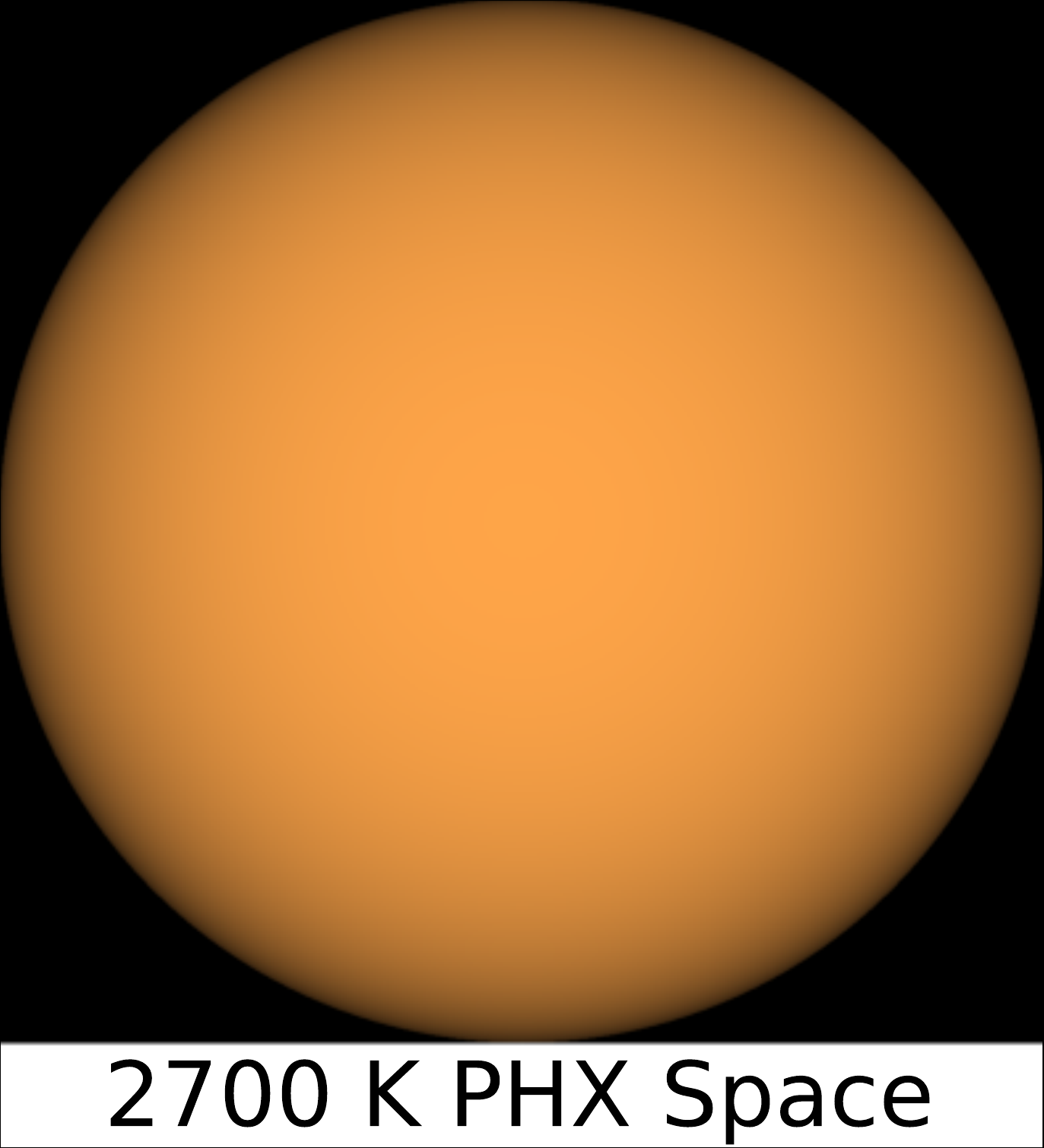}\\
\includegraphics[width=0.38\linewidth]{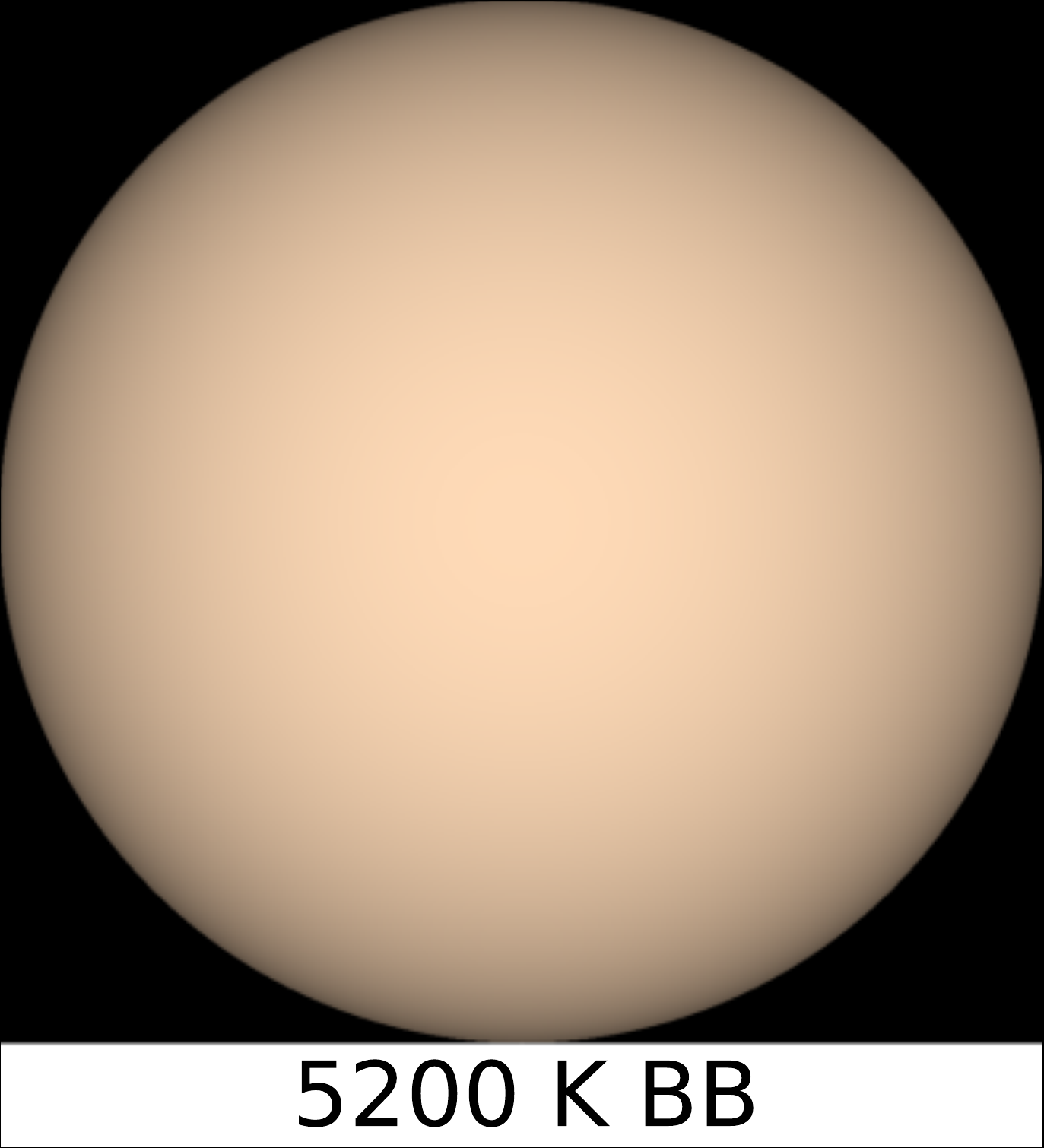}
\includegraphics[width=0.38\linewidth]{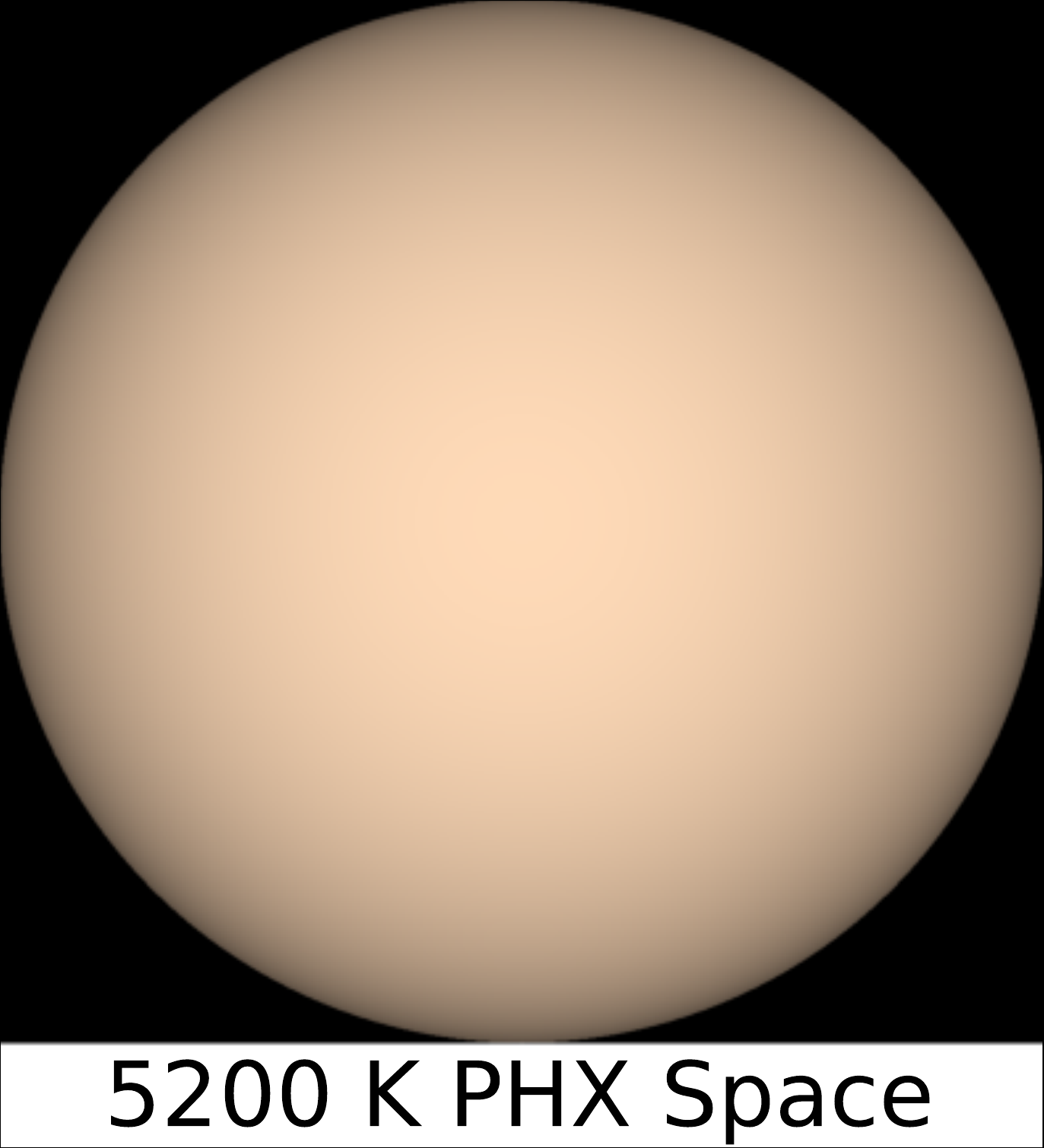}\\
\includegraphics[width=0.38\linewidth]{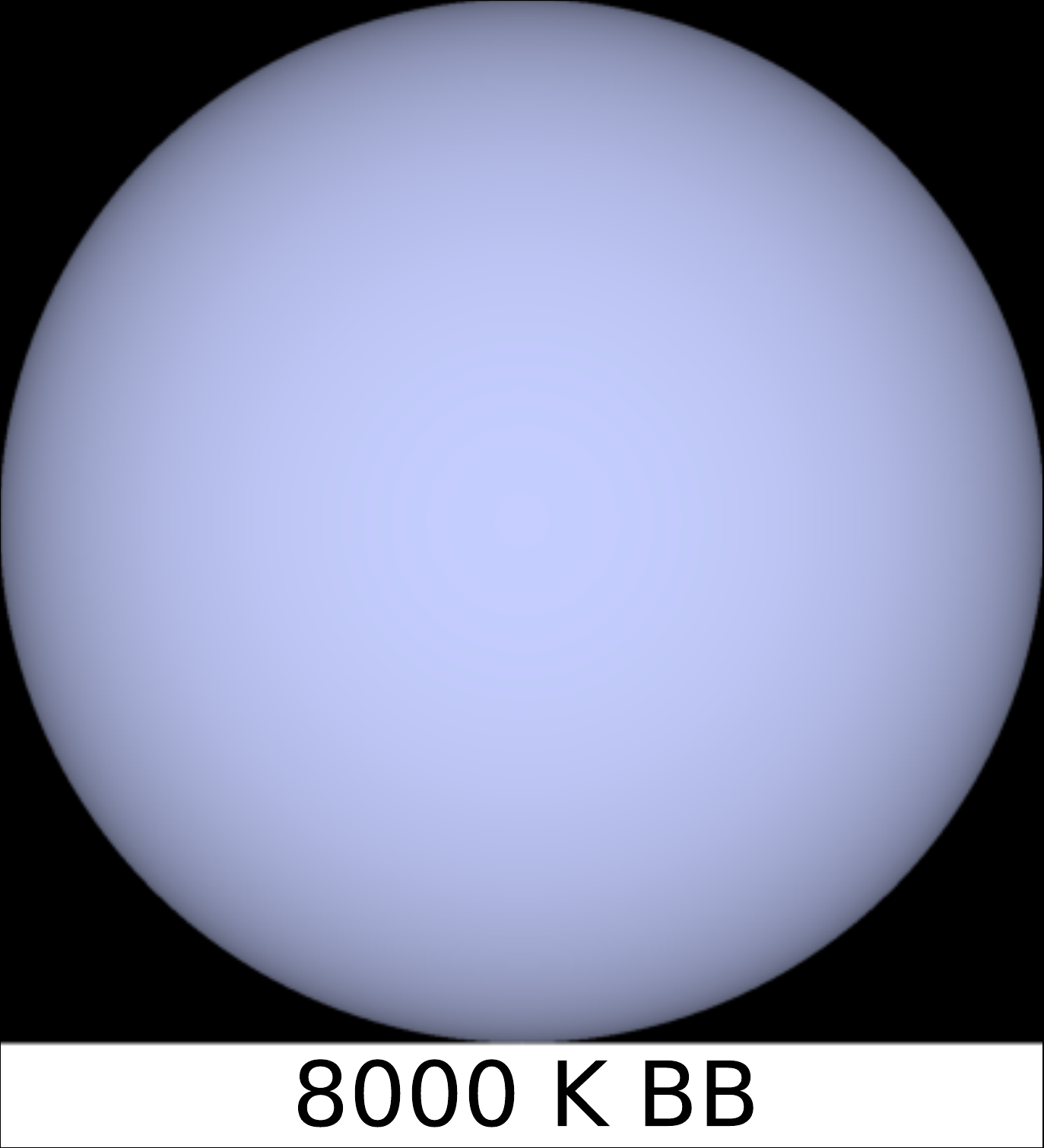}
\includegraphics[width=0.38\linewidth]{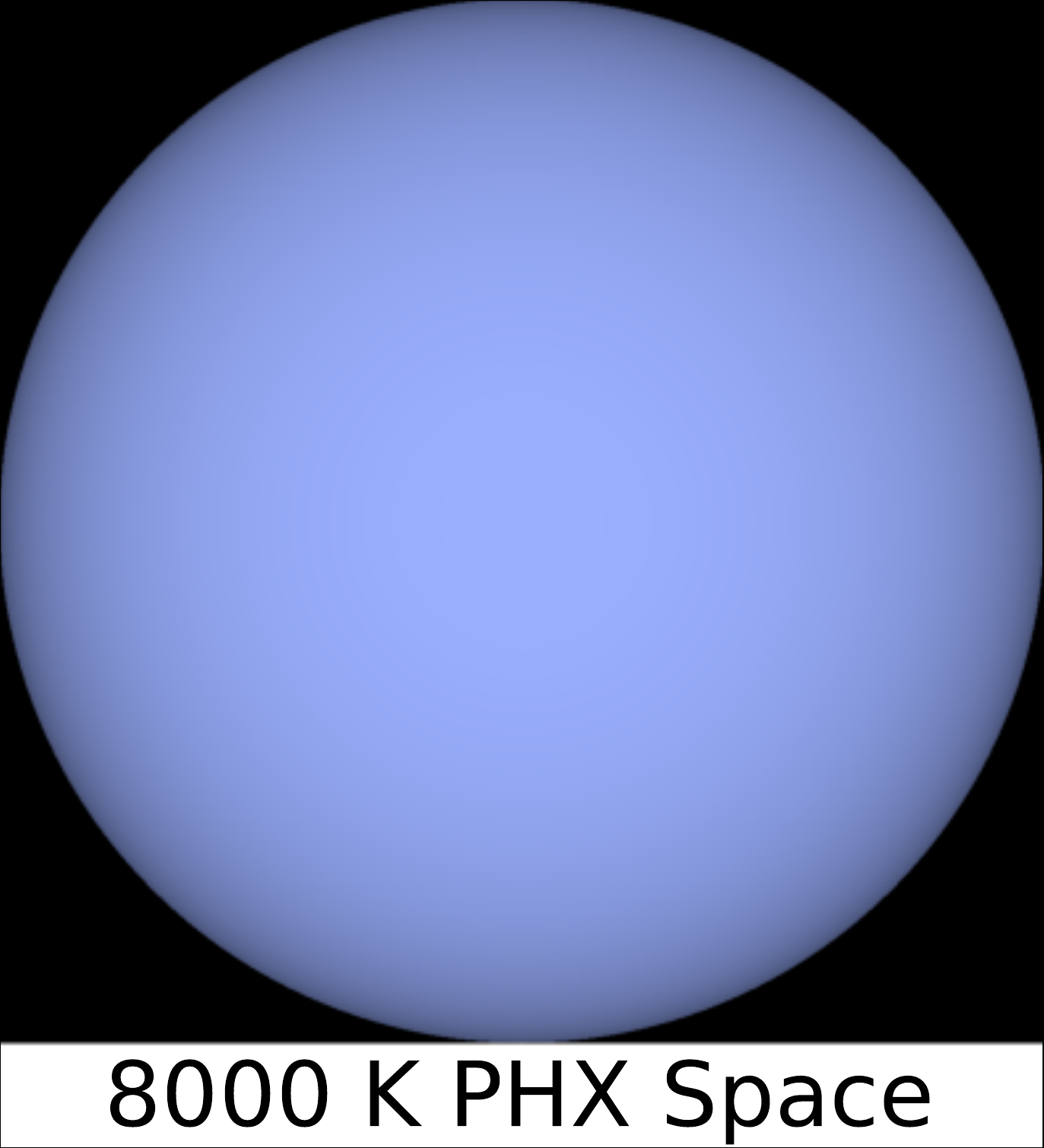}
\caption{Color representations of three stars with $T_{\rm eff}~=~2700$\,K (top), $T_{\rm eff}~=~5200$\,K (center), $T_{\rm eff}~=~8000$\,K (bottom). Colors in the left column are computed from a black body and colors in the right column from the PHOENIX spectra. For the PHOENIX models, we assumed solar metallicity ([{\rm Fe/H}]~=~0) and $\log(g)~=~5.0$, $\log(g)~=~4.5$, $\log(g)~=~4.0$, respectively. The corresponding color codes are listed in Table~\ref{tab:colors_Z0}. Quadratic limb darkening is modeled on top of the stellar disk. Perception of the colors (and their differences) depends on the monitor or print as well as on the individual vision abilities of the viewer.}
\label{fig:colors}
\end{figure}

The color differences between black bodies and stellar spectra mentioned above are illustrated in more detail in Fig.~\ref{fig:colors}. The left column shows stellar disks with limb darkening and colors based on the black-body approximation, and the right column refers to PHOENIX spectra as if observed from space.

In the upper row, which features a $2700\,$K star (corresponding to SpT M6.5), the black body disk appears redder than the PHOENIX star. We suppose that this reddish color derived from the black-body approximation could, at least partly, explain why the most early-type stars are called `red dwarfs'. The $5200\,$K black body radiator (left) in the center row shows almost no visible color difference to the PHOENIX star (SpT K1V, central panel). For $T_{\rm eff}~=~8000\,$K in the bottom row the black body disk looks lighter due to its slightly increased red values compared to the PHOENIX model disk. 

\begin{figure*}[t]
    \centering
    \includegraphics[width=1\linewidth]{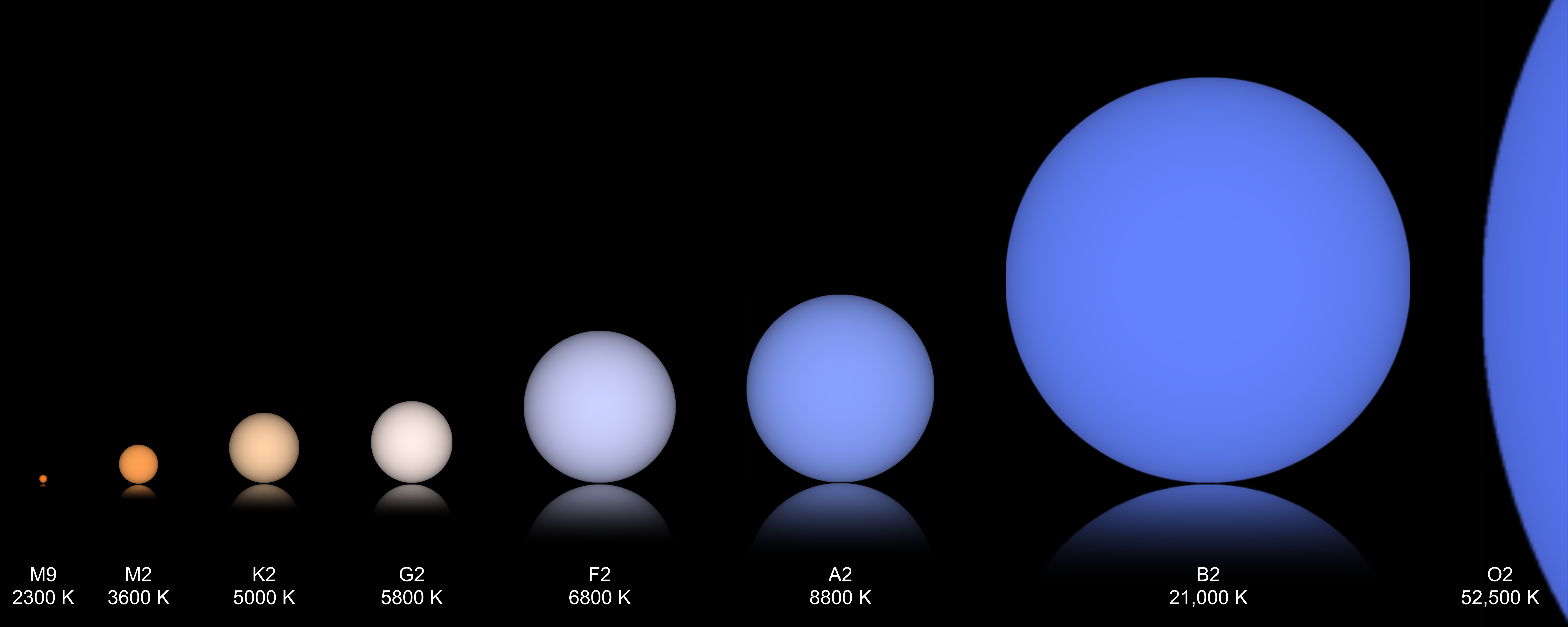}
    \caption{Selection of main-sequence stars with proper limb darkening, radii to scale and colors as computed in this report. All stars are of luminosity class V. From a human-perception point of view the plausibility of such an illustration is limited because the hottest stars (spectral type O) can be millions or billions times more luminous than the coolest stars of (spectral type M). We do not take into account the effect of brightness.}
    \label{fig:menagerie}
\end{figure*}

The menagerie of model main-sequence stars in Fig.~\ref{fig:menagerie} shows a summary of our astrophysical model. Colors are based on the PHOENIX models as if seen from space, the stellar disk is modeled using quadratic limb darkening, and stellar radii are scaled according to \citet{2018MNRAS.479.5491E} for main-sequence stars of the respective spectral types.

\section{Discussion}

Our color codes carry information about the chromaticity (hue and saturation) of stars but not about their brightness (see Fig.~\ref{fig:menagerie}). If we were to include stellar brightness as well, we would need to invoke the stellar luminosity and distance from Earth to compute an apparent brightness. The stellar luminosity could, in principle, be obtained from stellar evolution models or mass-luminosity scaling relations for stars of a given mass. The model spectra, however, do not make any assumptions of the stellar mass and so we would also have to assume a given age or evolutionary phase to match the resulting effective temperature and log($g$) between stellar evolution models and the synthetic spectra.

Our illustrations of the stellar disks with limb darkening consider an achromatic radial intensity profile. Chromatic effects have been observed in the sun, for example by \citet{1987SoPh..110..139N}. Strictly speaking, due to the wavelength dependence of stellar limb darkening, we should also expect a chromatic variation along the apparent distance from the stellar disk center. That said, the color codes that we calculate are derived from the model spectra directly and independent from our graphical illustrations.

The CMFs that we used came with a resolution of 1\,{\AA} thereby defining a limit to the accuracy of our calculations. We tested the effect of the resolution of the CMFs by decreasing their resolution artificially by a factor of two. We found that the black-body RGB color codes of some stars changed on the permille level, certainly below the level of perception for the human eye. For the PHOENIX spectra the situation was more complex. For $T_{\rm eff}>3000$\,K changes in the RGB color codes were $<1\,\%$, and therefore unlikely to be visible to most people. For even lower temperatures and down to the coolest M dwarfs with $T_{\rm eff}=2300$\,K, variations between our nominal results and the degraded resolution were always below $1\,\%$ for main-sequence stars. Only for the coolest stars with extremely low surface gravities of $\log(g)=0$, variations in the G color channel were more significant, up to 2.5\,\% for $2600\,{\rm K}~\leq~T_{\rm eff}~\leq~3000\,{\rm K}$ and up to 8\,\% for $2300\,{\rm K}~\leq~T_{\rm eff}~\leq~2600\,{\rm K}$. The R and B channels varied on a percent level for these cool and low-$g$ giants. In brief, the resolution of the CMFs is sufficient for the main-sequence stars to produce digital color codes exact to one permille for $T_{\rm eff}~>~3000$\,K and exact to one percent for $T_{\rm eff}~<~3000$\,K.

We also examined the impact of stellar rotation on the apparent colors. Stellar rotation leads to rotational broadening of the spectral absorption lines, which may or may not have a visible effect. We tested this by applying rotational broadening to a synthetic ATLAS spectrum \citep{1993ASPC...44...87K,2004MSAIS...5...93S} with $T_\mathrm{eff}=50,000\,$K and $\log(g) = 5.0$ with a corresponding linear limb darkening coefficient of $a~=~0.1680$ \citep{2019RNAAS...3...17C}. We assumed a rotational velocity of $v\sin(i) = 200\,$km$/$s to maximize the effect of a rapidly rotating early-type star \citep{2002ApJ...573..359A}. After applying the rotational broadening using the freely available {\tt rotBroad()}\footnote{\href{https://pyastronomy.readthedocs.io/en/latest/pyaslDoc/aslDoc/rotBroad.html}{https://pyastronomy.readthedocs.io/}} python function from the {\tt PyAstronomy} module and extracting the color codes from the spectrum, we found that the Hex color code remains unaffected and the RGB color code was only affected at the 7th decimal place. In summary, stellar rotation does not have a visible effect on the perceived colors of stars.

CMFs are usually derived from a series of experiments with test persons. Different test series derive different CMFs, although results are very similar \citep{North1993}. Thus we expect that choosing a different set of CMFs would result in virtually the same digital color codes of main-sequence stars. Irrespective of the color codes that we provide, the perception of the colors and of the apparent color differences depends heavily on the monitor or printer used to display the star images. The color vision abilities of the viewer also affect the hue, that is to say, the color appearance of any light source.

Another potentially relevant effect on color perception is the Purkinje effect \citep{graff2013grundriss}, which makes reds appear darker and blues brighter at low brightness. Our color codes are determined as if the stars were observed under standard illumination, which is a somewhat artificial experimental setup for a hypothetical by-eye observation from space, admittedly. In a real space-based observation of a star with the unaided human eye, the Purkinje effect could lead to a slight difference in the perceived colors compared to our color codes.

In this paper, we restrict ourselves to the apparent colors of stars as if seen from space. For an observer on Earth, the colors would look slightly different due to several effects in the Earth's atmosphere such as scattering of the light in the atmosphere, scintillation, Rayleigh scattering, and the presence of carbon black dust and other small particles \citep[Chapter~19]{1991cahe.book.....S}. 
That said, we did a small experiment of simulating the transmission of stellar light through the Earth's atmosphere using data from the HITRAN\footnote{\href{https://hitran.org/}{https://hitran.org}} database \citep{2017JQSRT.203....3G}. Every molecule and isotopologue listed in the database, except for the CO$_2$ isotopologues $^{18}$C$^{13}$C$^{17}$O and $^{13}$C$^{17}$O$_2$, was used in the generation of the absorption spectrum. The resulting transmission spectrum was binned to the same wavelength grid as the CMFs with a fixed wavelength step size of 1\,\AA. The transmission spectrum was normalized to a maximum value of one and then multiplied with the respective PHOENIX spectrum. The transmission spectrum for the Earth's atmosphere was normalized to the maximum value in the wavelength range of the model spectra. A different normalization would result in similar colors, since the absorption of light in the Earth's atmosphere only has a small impact $(\sim0.1\%)$ on the RGB color code. As a result, we found that the effects from transmission on the stellar RGB color codes are less than about 1\,\%. The other atmospheric effects are probably more significant.

The digital color codes that we computed for black bodies can safely be used as representative for white dwarfs of spectral type DC. Spectra of DC white dwarfs are very close to black body radiators with essentially no significant absorption lines \citep{Greenstein1958,1983ApJ...275..240W}. This has some remarkable implications for DC white dwarfs with cooling ages of billions of years, at which point their effective temperatures drop below 5000\,K. The look of these white dwarfs is a light orange. Their color is similar to the star of SpT K2 in Fig.~\ref{fig:menagerie}. The stellar remnant of the recently discovered transiting exoplanet candidate WD\,1856$+$534\,b is a topical example \citep{Vanderburg2020}. With $T_{\rm eff}=4710\,\pm60$\,K this DC white dwarf would really have a pale orange look to the human eye.


\section{Conclusions}

We use standard color matching functions representing the color perception of the human eye to determine the colors of stars from synthetic spectra. We find that colors derived from realistic stellar spectra differ substantially from colors based on a black body approximation.

In particular, the coolest ($T_{\rm eff}=2300$\,K) and thus most late-\linebreak type (SpT M9.5V) main-sequence stars look orange to the human eye. The digital linear RGB color code of a solar metallicity M9.5V dwarf reads (1.0, 0.491, 0.144) (Table~\ref{tab:SpT_colors_2}). We suppose that part of the explanation for why these stars are often called red dwarfs is in the fact that a black body spectrum yields a much redder color, namely RGB = (1.0, 0.409, 0.078) (Table~\ref{tab:colors_bb}). The colors of stars of SpT K1V - K0V (with $T_{\rm eff}=5200$\,K or 5300\,K, respectively) are very similar to those of a black body with the same effective temperature. Early-type stars of SpT F, A, B, and O look bluer than their black-body counterparts because synthetic spectra yield lower red values in the resulting linear RGB color codes.

We find that the transmission of star light through the Earth's atmosphere has a negligible effect on color perception, with color differences on the level of a few permille in the corresponding RGB color codes.

Our ad-hoc parameterization of stellar limb darkening with the quadratic limb darkening law yields realistic, plastic illustrations of main-sequence stars with correct colors, achromatic limb darkening, and physical stellar radii to scale. To our knowledge, this is the first digital representation of main-sequence stars with consistent and correct colors based on astrophysical modeling. Our computer code, {\tt Spec2Col.py}, is freely available and can be used to compute digital colors codes for any type of input spectrum. Images of 107 stars with spectral types ranging from M9.5V to O1V as per Table~\ref{tab:SpT_colors_2} including limb darkening are available as PDF with the online version of this paper.

\section*{Acknowledgments}
The authors appreciate the comments on the manuscript from an anonymous referee. The authors also thank Christian Hill for providing them with the {\tt color\_system.py} code, Tim-Oliver Husser for help in accessing the PHOENIX files, and Felix Mackebrandt for comments on a manuscript draft. This work made use of NASA's ADS Bibliographic Services. RH is supported by the \fundingAgency{German Aerospace Agency (Deutsches Zentrum f{\"u}r Luft- und Raumfahrt)} under PLATO Data Center grant \fundingNumber{50OO1501}.

%

\subsection*{Conflict of interest}

The authors declare no potential conflict of interests.

\section*{Supporting information}

The data underlying this article are available on Zenodo at \href{https://doi.org/10.5281/zenodo.4090873}{https://doi.org/10.5281/zenodo.4090873}.

\appendix

\clearpage

\begin{table*}
\footnotesize
\caption{Linear RGB and Hex color codes of black bodies. The full table is available online (see Supporting information).}
\def\arraystretch{1.2}
\label{tab:colors_bb}
\centering
\begin{tabular}{c | c c }
\hline\hline
 $T_{\mathrm{eff}}$ & RGB & Hex \\\hline
2300 & 1.0,0.409,0.078 & \#ff6813 \\ 
2400 & 1.0,0.432,0.093 & \#ff6e17 \\ 
2500 & 1.0,0.455,0.109 & \#ff731b \\ 
2600 & 1.0,0.476,0.126 & \#ff7920 \\ 
2700 & 1.0,0.497,0.144 & \#ff7e24 \\ 
2800 & 1.0,0.518,0.163 & \#ff8429 \\ 
2900 & 1.0,0.537,0.182 & \#ff892e \\ 
3000 & 1.0,0.557,0.202 & \#ff8d33 \\ 
3100 & 1.0,0.575,0.223 & \#ff9238 \\ 
3200 & 1.0,0.593,0.244 & \#ff973e \\ 
3300 & 1.0,0.611,0.266 & \#ff9b43 \\ 
3400 & 1.0,0.627,0.289 & \#ff9f49 \\ 
3500 & 1.0,0.644,0.311 & \#ffa44f \\ 
... & ... & ...  \\
\hline
\end{tabular}
\end{table*}

\begin{table*}
\footnotesize
\caption{Linear RGB and Hex color codes of stars with solar metallicity ($[{\mathrm{Fe/H}}]~=~0$) as seen from space, computed using PHOENIX model spectra. The full table is available online (see Supporting information).}
\def\arraystretch{1.2}
\label{tab:colors_Z0}
\centering
\begin{tabular}{c c c | c c }
\hline\hline
 $T_{\mathrm{eff}}$ & $\log(g)$ & $[{\mathrm{Fe/H}}]$ & RGB & Hex \\\hline
2300 & 3.0 & 0.0 & 1.0,0.615,0.292 & \#ff9c4a \\ 
2300 & 3.5 & 0.0 & 1.0,0.571,0.25 & \#ff913f \\ 
2300 & 4.0 & 0.0 & 1.0,0.539,0.211 & \#ff8935 \\ 
2300 & 4.5 & 0.0 & 1.0,0.507,0.175 & \#ff812c \\ 
2300 & 5.0 & 0.0 & 1.0,0.491,0.144 & \#ff7d24 \\ 
2300 & 5.5 & 0.0 & 1.0,0.435,0.098 & \#ff6e19 \\ 
2300 & 6.0 & 0.0 & 1.0,0.393,0.076 & \#ff6413 \\ 
2400 & 3.0 & 0.0 & 1.0,0.66,0.313 & \#ffa84f \\ 
2400 & 3.5 & 0.0 & 1.0,0.653,0.309 & \#ffa64e \\ 
2400 & 4.0 & 0.0 & 1.0,0.604,0.266 & \#ff9a43 \\ 
2400 & 4.5 & 0.0 & 1.0,0.549,0.212 & \#ff8b36 \\ 
2400 & 5.0 & 0.0 & 1.0,0.518,0.179 & \#ff842d \\ 
2400 & 5.5 & 0.0 & 1.0,0.484,0.137 & \#ff7b23 \\ 
2400 & 6.0 & 0.0 & 1.0,0.434,0.098 & \#ff6e19 \\
... & ... & ... & ... & ... \\
\hline
\end{tabular}
\end{table*}

\begin{table*}
\footnotesize
\caption{Linear RGB and Hex color codes of stars with $[{\mathrm{Fe/H}}]~=~-1$ as seen from space, computed using PHOENIX model spectra. The full table is available online (see Supporting information).}
\def\arraystretch{1.2}
\label{tab:colors_Z-1}
\centering
\begin{tabular}{c c c | c c }
\hline\hline
 $T_{\mathrm{eff}}$ & $\log(g)$ & $[{\mathrm{Fe/H}}]$ & RGB & Hex \\\hline
2300 & 3.0 & -1.0 & 1.0,0.752,0.303 & \#ffbf4d \\ 
2300 & 3.5 & -1.0 & 1.0,0.637,0.213 & \#ffa236 \\ 
2300 & 4.0 & -1.0 & 1.0,0.559,0.151 & \#ff8e26 \\ 
2300 & 4.5 & -1.0 & 1.0,0.493,0.103 & \#ff7d1a \\ 
2300 & 5.0 & -1.0 & 1.0,0.469,0.086 & \#ff7715 \\ 
2300 & 5.5 & -1.0 & 1.0,0.425,0.059 & \#ff6c0f \\ 
2300 & 6.0 & -1.0 & 1.0,0.392,0.045 & \#ff640b \\ 
2400 & 3.0 & -1.0 & 1.0,0.774,0.312 & \#ffc54f \\ 
2400 & 3.5 & -1.0 & 1.0,0.671,0.232 & \#ffab3b \\ 
2400 & 4.0 & -1.0 & 1.0,0.604,0.183 & \#ff992e \\ 
2400 & 4.5 & -1.0 & 1.0,0.517,0.123 & \#ff831f \\ 
2400 & 5.0 & -1.0 & 1.0,0.468,0.09 & \#ff7717 \\ 
2400 & 5.5 & -1.0 & 1.0,0.426,0.063 & \#ff6c0f \\ 
2400 & 6.0 & -1.0 & 1.0,0.406,0.052 & \#ff670d \\ 
... & ... & ... & ... & ...   \\
\hline
\end{tabular}
\end{table*}

\begin{table*}
\footnotesize
\caption{Linear RGB and Hex color codes of stars with $[{\mathrm{Fe/H}}]~=~-2$ as seen from space, computed using PHOENIX model spectra. The full table is available online (see Supporting information).}
\def\arraystretch{1.2}
\label{tab:colors_Z-2}
\centering
\begin{tabular}{c c c | c c }
\hline\hline
 $T_{\mathrm{eff}}$ & $\log(g)$ & $[{\mathrm{Fe/H}}]$ & RGB & Hex \\\hline
2300 & 3.0 & -2.0 & 1.0,0.525,0.082 & \#ff8514 \\ 
2300 & 3.5 & -2.0 & 1.0,0.48,0.067 & \#ff7a10 \\ 
2300 & 4.0 & -2.0 & 1.0,0.443,0.057 & \#ff710e \\ 
2300 & 4.5 & -2.0 & 1.0,0.422,0.054 & \#ff6b0d \\ 
2300 & 5.0 & -2.0 & 1.0,0.41,0.052 & \#ff680d \\ 
2300 & 5.5 & -2.0 & 1.0,0.394,0.051 & \#ff640d \\ 
2300 & 6.0 & -2.0 & 1.0,0.382,0.055 & \#ff610d \\ 
2400 & 3.0 & -2.0 & 1.0,0.527,0.09 & \#ff8617 \\ 
2400 & 3.5 & -2.0 & 1.0,0.506,0.082 & \#ff8115 \\ 
2400 & 4.0 & -2.0 & 1.0,0.468,0.07 & \#ff7711 \\ 
2400 & 4.5 & -2.0 & 1.0,0.437,0.064 & \#ff6f10 \\ 
2400 & 5.0 & -2.0 & 1.0,0.417,0.062 & \#ff6a0f \\ 
2400 & 5.5 & -2.0 & 1.0,0.406,0.061 & \#ff670f \\ 
2400 & 6.0 & -2.0 & 1.0,0.394,0.065 & \#ff6410 \\ 
... & ... & ... & ... & ...  \\
\hline
\end{tabular}
\end{table*}

\begin{table*}
\footnotesize
\caption{Linear RGB and Hex color codes of solar metallicity main sequence stars for spectral types (SpT) M9.5V - O1V.}
\def\arraystretch{1.2}
\label{tab:SpT_colors}
\centering
\begin{tabular}{c c c c c c}
\hline\hline
 SpT & $T_{\mathrm{eff}}$ & $\log(g)$ & RGB & Hex \\  \hline
M9.5V & 2300 & 5.0 & 1.0,0.491,0.144 & \#ff7d24 & \LARGE \textcolor[rgb]{1.0,0.491,0.144}{$\bullet$} \\
M9V & 2400 & 5.0 & 1.0,0.518,0.179 & \#ff842d & \LARGE \textcolor[rgb]{1.0,0.518,0.179}{$\bullet$} \\
M8V & 2500 & 5.0 & 1.0,0.542,0.202 & \#ff8a33 & \LARGE \textcolor[rgb]{1.0,0.542,0.202}{$\bullet$} \\
M7.5V & 2600 & 5.0 & 1.0,0.607,0.255 & \#ff9a41 & \LARGE \textcolor[rgb]{1.0,0.607,0.255}{$\bullet$} \\
M6.5V & 2700 & 5.0 & 1.0,0.648,0.286 & \#ffa548 & \LARGE \textcolor[rgb]{1.0,0.648,0.286}{$\bullet$} \\
M6V & 2800 & 5.0 & 1.0,0.649,0.285 & \#ffa548 & \LARGE \textcolor[rgb]{1.0,0.649,0.285}{$\bullet$} \\
M6V & 2900 & 5.0 & 1.0,0.644,0.285 & \#ffa448 & \LARGE \textcolor[rgb]{1.0,0.644,0.285}{$\bullet$} \\
M5.5V & 3000 & 5.0 & 1.0,0.641,0.289 & \#ffa349 & \LARGE \textcolor[rgb]{1.0,0.641,0.289}{$\bullet$} \\
M4.5V & 3100 & 5.0 & 1.0,0.638,0.293 & \#ffa24a & \LARGE \textcolor[rgb]{1.0,0.638,0.293}{$\bullet$} \\
M4V & 3200 & 5.0 & 1.0,0.638,0.3 & \#ffa24c & \LARGE \textcolor[rgb]{1.0,0.638,0.3}{$\bullet$} \\ 
M3.5V & 3300 & 5.0 & 1.0,0.638,0.308 & \#ffa24e & \LARGE \textcolor[rgb]{1.0,0.638,0.308}{$\bullet$} \\
M3V & 3400 & 5.0 & 1.0,0.638,0.315 & \#ffa250 & \LARGE \textcolor[rgb]{1.0,0.638,0.315}{$\bullet$} \\
M2.5V & 3500 & 5.0 & 1.0,0.637,0.322 & \#ffa251 & \LARGE \textcolor[rgb]{1.0,0.637,0.322}{$\bullet$} \\
M2V & 3600 & 5.0 & 1.0,0.635,0.327 & \#ffa153 & \LARGE \textcolor[rgb]{1.0,0.635,0.327}{$\bullet$} \\
M1V & 3700 & 4.5 & 1.0,0.637,0.34 & \#ffa256 & \LARGE \textcolor[rgb]{1.0,0.637,0.34}{$\bullet$} \\
M0.5V & 3800 & 4.5 & 1.0,0.635,0.346 & \#ffa158 & \LARGE \textcolor[rgb]{1.0,0.635,0.346}{$\bullet$} \\ 
M0V & 3900 & 4.5 & 1.0,0.636,0.354 & \#ffa25a & \LARGE \textcolor[rgb]{1.0,0.636,0.354}{$\bullet$} \\
K8V & 4000 & 4.5 & 1.0,0.641,0.369 & \#ffa35e & \LARGE \textcolor[rgb]{1.0,0.641,0.369}{$\bullet$} \\
K7V & 4100 & 4.5 & 1.0,0.65,0.389 & \#ffa563 & \LARGE \textcolor[rgb]{1.0,0.65,0.389}{$\bullet$} \\ 
K6.5V & 4200 & 4.5 & 1.0,0.662,0.411 & \#ffa868 & \LARGE \textcolor[rgb]{1.0,0.662,0.411}{$\bullet$} \\ 
K5.5V & 4300 & 4.5 & 1.0,0.677,0.439 & \#ffac6f & \LARGE \textcolor[rgb]{1.0,0.677,0.439}{$\bullet$} \\
K5V & 4400 & 4.5 & 1.0,0.696,0.47 & \#ffb177 & \LARGE \textcolor[rgb]{1.0,0.696,0.47}{$\bullet$} \\
K4.5V & 4500 & 4.5 & 1.0,0.717,0.501 & \#ffb67f & \LARGE \textcolor[rgb]{1.0,0.717,0.501}{$\bullet$} \\
K4V & 4600 & 4.5 & 1.0,0.739,0.533 & \#ffbc87 & \LARGE \textcolor[rgb]{1.0,0.739,0.533}{$\bullet$} \\
K3.5V & 4700 & 4.5 & 1.0,0.761,0.565 & \#ffc18f & \LARGE \textcolor[rgb]{1.0,0.761,0.565}{$\bullet$} \\
K3V & 4800 & 4.5 & 1.0,0.781,0.595 & \#ffc797 & \LARGE \textcolor[rgb]{1.0,0.781,0.595}{$\bullet$} \\
K3V & 4900 & 4.5 & 1.0,0.802,0.626 & \#ffcc9f & \LARGE \textcolor[rgb]{1.0,0.802,0.626}{$\bullet$} \\
K2.5V & 5000 & 4.5 & 1.0,0.821,0.657 & \#ffd1a7 & \LARGE \textcolor[rgb]{1.0,0.821,0.657}{$\bullet$} \\
K1.5V & 5100 & 4.5 & 1.0,0.84,0.691 & \#ffd6b0 & \LARGE \textcolor[rgb]{1.0,0.84,0.691}{$\bullet$} \\
K1V & 5200 & 4.5 & 1.0,0.857,0.722 & \#ffdab8 & \LARGE \textcolor[rgb]{1.0,0.857,0.722}{$\bullet$} \\ 
\hline
\end{tabular}
\end{table*}

\setcounter{table}{4}

\begin{table*}
\footnotesize
\caption{{\textit{(continued)}}}
\def\arraystretch{1.2}
\label{tab:SpT_colors_2}
\centering
\begin{tabular}{c c c c c c}
\hline\hline
SpT & $T_{\mathrm{eff}}$ & $\log(g)$ & RGB & Hex \\  \hline
K0V & 5300 & 4.5 & 1.0,0.872,0.753 & \#ffdec0 & \LARGE \textcolor[rgb]{1.0,0.872,0.753}{$\bullet$} \\
G9V & 5400 & 4.5 & 1.0,0.886,0.783 & \#ffe1c7 & \LARGE \textcolor[rgb]{1.0,0.886,0.783}{$\bullet$} \\
G8V & 5500 & 4.5 & 1.0,0.898,0.813 & \#ffe5cf & \LARGE \textcolor[rgb]{1.0,0.898,0.813}{$\bullet$} \\
G6V & 5600 & 4.5 & 1.0,0.91,0.845 & \#ffe8d7 & \LARGE \textcolor[rgb]{1.0,0.91,0.845}{$\bullet$} \\
G4V & 5700 & 4.5 & 1.0,0.922,0.878 & \#ffebdf & \LARGE \textcolor[rgb]{1.0,0.922,0.878}{$\bullet$} \\
G2V & 5800 & 4.5 & 1.0,0.931,0.905 & \#ffede6 & \LARGE \textcolor[rgb]{1.0,0.931,0.905}{$\bullet$} \\
G1V & 5900 & 4.5 & 1.0,0.94,0.931 & \#ffefed & \LARGE \textcolor[rgb]{1.0,0.94,0.931}{$\bullet$} \\
F9.5V & 6000 & 4.5 & 1.0,0.951,0.967 & \#fff2f6 & \LARGE \textcolor[rgb]{1.0,0.951,0.967}{$\bullet$} \\
F9V & 6100 & 4.5 & 1.0,0.96,0.998 & \#fff4fe & \LARGE \textcolor[rgb]{1.0,0.96,0.998}{$\bullet$} \\
F8V & 6200 & 4.0 & 0.955,0.931,1.0 & \#f3edff & \LARGE \textcolor[rgb]{0.955,0.931,1.0}{$\bullet$} \\ 
F6V & 6300 & 4.0 & 0.922,0.908,1.0 & \#ebe7ff & \LARGE \textcolor[rgb]{0.922,0.908,1.0}{$\bullet$} \\
F6V & 6400 & 4.0 & 0.896,0.891,1.0 & \#e4e3ff & \LARGE \textcolor[rgb]{0.896,0.891,1.0}{$\bullet$} \\
F5V & 6500 & 4.0 & 0.869,0.871,1.0 & \#dddeff & \LARGE \textcolor[rgb]{0.869,0.871,1.0}{$\bullet$} \\
F4V & 6600 & 4.0 & 0.844,0.855,1.0 & \#d7d9ff & \LARGE \textcolor[rgb]{0.844,0.855,1.0}{$\bullet$} \\
F3V & 6700 & 4.0 & 0.823,0.84,1.0 & \#d1d6ff & \LARGE \textcolor[rgb]{0.823,0.84,1.0}{$\bullet$} \\
F2V & 6800 & 4.0 & 0.802,0.826,1.0 & \#ccd2ff & \LARGE \textcolor[rgb]{0.802,0.826,1.0}{$\bullet$} \\
F2V & 6900 & 4.0 & 0.782,0.812,1.0 & \#c7cfff & \LARGE \textcolor[rgb]{0.782,0.812,1.0}{$\bullet$} \\
F1V & 7000 & 4.0 & 0.763,0.799,1.0 & \#c2cbff & \LARGE \textcolor[rgb]{0.763,0.799,1.0}{$\bullet$} \\
F0V & 7200 & 4.0 & 0.725,0.773,1.0 & \#b8c5ff & \LARGE \textcolor[rgb]{0.725,0.773,1.0}{$\bullet$} \\
A9V & 7400 & 4.0 & 0.692,0.75,1.0 & \#b0bfff & \LARGE \textcolor[rgb]{0.692,0.75,1.0}{$\bullet$} \\ 
A8V & 7600 & 4.0 & 0.674,0.738,1.0 & \#abbcff & \LARGE \textcolor[rgb]{0.674,0.738,1.0}{$\bullet$} \\
A7V & 7800 & 4.0 & 0.636,0.712,1.0 & \#a2b5ff & \LARGE \textcolor[rgb]{0.636,0.712,1.0}{$\bullet$} \\
A6V & 8000 & 4.0 & 0.606,0.69,1.0 & \#9ab0ff & \LARGE \textcolor[rgb]{0.606,0.69,1.0}{$\bullet$} \\
A4V & 8200 & 4.0 & 0.579,0.67,1.0 & \#93aaff & \LARGE \textcolor[rgb]{0.579,0.67,1.0}{$\bullet$} \\
A4V & 8400 & 4.0 & 0.556,0.652,1.0 & \#8da6ff & \LARGE \textcolor[rgb]{0.556,0.652,1.0}{$\bullet$} \\
A3V & 8600 & 4.0 & 0.546,0.645,1.0 & \#8ba4ff & \LARGE \textcolor[rgb]{0.546,0.645,1.0}{$\bullet$} \\
A2V & 8800 & 4.0 & 0.531,0.634,1.0 & \#87a1ff & \LARGE \textcolor[rgb]{0.531,0.634,1.0}{$\bullet$} \\
A2V & 9000 & 4.0 & 0.519,0.624,1.0 & \#849fff & \LARGE \textcolor[rgb]{0.519,0.624,1.0}{$\bullet$} \\
A1V & 9200 & 4.0 & 0.508,0.616,1.0 & \#819dff & \LARGE \textcolor[rgb]{0.508,0.616,1.0}{$\bullet$} \\
A1V & 9400 & 4.0 & 0.498,0.608,1.0 & \#7f9bff & \LARGE \textcolor[rgb]{0.498,0.608,1.0}{$\bullet$} \\
A0V & 9600 & 4.0 & 0.49,0.601,1.0 & \#7d99ff & \LARGE \textcolor[rgb]{0.49,0.601,1.0}{$\bullet$} \\
A0V & 9800 & 4.0 & 0.483,0.595,1.0 & \#7b97ff & \LARGE \textcolor[rgb]{0.483,0.595,1.0}{$\bullet$} \\
A0V & 10000 & 4.0 & 0.477,0.59,1.0 & \#7996ff & \LARGE \textcolor[rgb]{0.477,0.59,1.0}{$\bullet$} \\
B9.5V & 10200 & 4.0 & 0.472,0.586,1.0 & \#7895ff & \LARGE \textcolor[rgb]{0.472,0.586,1.0}{$\bullet$} \\
B9.5V & 10400 & 4.0 & 0.467,0.582,1.0 & \#7794ff & \LARGE \textcolor[rgb]{0.467,0.582,1.0 }{$\bullet$} \\
B9V & 10600 & 4.0 & 0.463,0.578,1.0 & \#7693ff & \LARGE \textcolor[rgb]{0.463,0.578,1.0}{$\bullet$} \\
B9V & 10800 & 4.0 & 0.459,0.575,1.0 & \#7592ff & \LARGE \textcolor[rgb]{0.459,0.575,1.0}{$\bullet$} \\
B9V & 11000 & 4.0 & 0.456,0.572,1.0 & \#7491ff & \LARGE \textcolor[rgb]{0.456,0.572,1.0}{$\bullet$} \\
B9V & 11200 & 4.0 & 0.453,0.57,1.0 & \#7391ff & \LARGE \textcolor[rgb]{0.453,0.57,1.0}{$\bullet$} \\
B9V & 11400 & 4.0 & 0.451,0.567,1.0 & \#7290ff & \LARGE \textcolor[rgb]{0.451,0.567,1.0}{$\bullet$} \\
B9V & 11600 & 4.0 & 0.45,0.566,1.0 & \#7290ff & \LARGE \textcolor[rgb]{0.45,0.566,1.0}{$\bullet$} \\
B8V & 11800 & 4.0 & 0.448,0.564,1.0 & \#728fff & \LARGE \textcolor[rgb]{0.448,0.564,1.0}{$\bullet$} \\
B8V & 12000 & 4.0 & 0.446,0.562,1.0 & \#718fff & \LARGE \textcolor[rgb]{0.446,0.562,1.0}{$\bullet$} \\
B8V & 12500 & 4.0 & 0.44,0.557,1.0 & \#708eff & \LARGE \textcolor[rgb]{0.44,0.557,1.0}{$\bullet$}\\ 
B8V & 13000 & 4.0 & 0.436,0.554,1.0 & \#6f8dff & \LARGE \textcolor[rgb]{0.436,0.554,1.0}{$\bullet$}\\ 
B7V & 13500 & 4.0 & 0.432,0.55,1.0 & \#6e8cff & \LARGE \textcolor[rgb]{0.432,0.55,1.0}{$\bullet$}\\ 
B7V & 14000 & 4.0 & 0.429,0.547,1.0 & \#6d8bff & \LARGE \textcolor[rgb]{0.429,0.547,1.0}{$\bullet$}\\ 
B6V & 14500 & 4.0 & 0.425,0.544,1.0 & \#6c8aff & \LARGE \textcolor[rgb]{0.425,0.544,1.0}{$\bullet$}\\ 
B6V & 15000 & 4.0 & 0.421,0.541,1.0 & \#6b89ff & \LARGE \textcolor[rgb]{0.421,0.541,1.0}{$\bullet$}\\ 
B5V & 16000 & 4.0 & 0.414,0.536,1.0 & \#6988ff & \LARGE \textcolor[rgb]{0.414,0.536,1.0}{$\bullet$} \\
\hline
\end{tabular}
\end{table*}

\setcounter{table}{4}

\begin{table*}
\footnotesize
\caption{{\textit{(continued)}}}
\def\arraystretch{1.2}
\label{tab:SpT_colors_2}
\centering
\begin{tabular}{c c c c c c}
\hline\hline
SpT & $T_{\mathrm{eff}}$ & $\log(g)$ & RGB & Hex \\  \hline
B3V & 17000 & 4.0 & 0.408,0.532,1.0 & \#6887ff & \LARGE \textcolor[rgb]{0.408,0.532,1.0}{$\bullet$} \\ 
B2.5V & 18000 & 4.0 & 0.403,0.527,1.0 & \#6686ff & \LARGE \textcolor[rgb]{0.403,0.527,1.0}{$\bullet$} \\ 
B2.5V & 19000 & 4.0 & 0.398,0.524,1.0 & \#6585ff & \LARGE \textcolor[rgb]{0.398,0.524,1.0}{$\bullet$} \\ 
B2V & 20000 & 4.0 & 0.394,0.52,1.0 & \#6484ff & \LARGE \textcolor[rgb]{0.394,0.52,1.0}{$\bullet$} \\ 
B2V & 21000 & 4.0 & 0.39,0.517,1.0 & \#6383ff & \LARGE \textcolor[rgb]{0.39,0.517,1.0}{$\bullet$} \\ 
B2V & 22000 & 4.0 & 0.387,0.514,1.0 & \#6283ff & \LARGE \textcolor[rgb]{0.387,0.514,1.0}{$\bullet$} \\ 
B1.5V & 23000 & 4.0 & 0.384,0.512,1.0 & \#6182ff & \LARGE \textcolor[rgb]{0.384,0.512,1.0}{$\bullet$} \\ 
B1.5V & 24000 & 4.0 & 0.381,0.509,1.0 & \#6181ff & \LARGE \textcolor[rgb]{0.381,0.509,1.0}{$\bullet$} \\ 
B1.5V & 25000 & 4.0 & 0.379,0.507,1.0 & \#6081ff & \LARGE \textcolor[rgb]{0.379,0.507,1.0}{$\bullet$} \\ 
B1V & 26000 & 4.0 & 0.376,0.505,1.0 & \#5f80ff & \LARGE \textcolor[rgb]{0.376,0.505,1.0}{$\bullet$} \\
B1V & 27000 & 4.0 & 0.373,0.503,1.0 & \#5f80ff & \LARGE \textcolor[rgb]{0.373,0.503,1.0}{$\bullet$} \\ 
B1V & 27500 & 4.0 & 0.371,0.501,1.0 & \#5e7fff & \LARGE \textcolor[rgb]{0.371,0.501,1.0}{$\bullet$} \\ 
B0.5V & 28000 & 4.0 & 0.371,0.5,1.0 & \#5e7fff & \LARGE \textcolor[rgb]{0.371,0.5,1.0}{$\bullet$} \\ 
B0.5V & 29000 & 4.0 & 0.368,0.498,1.0 & \#5d7fff & \LARGE \textcolor[rgb]{0.368,0.498,1.0}{$\bullet$} \\ 
B0.5V & 30000 & 4.0 & 0.366,0.496,1.0 & \#5d7eff & \LARGE \textcolor[rgb]{0.366,0.496,1.0}{$\bullet$} \\ 
O9.5V & 32500 & 4.0 & 0.361,0.491,1.0 & \#5c7dff & \LARGE \textcolor[rgb]{0.361,0.491,1.0}{$\bullet$} \\ 
O8V & 35000 & 4.0 & 0.357,0.487,1.0 & \#5b7cff & \LARGE \textcolor[rgb]{0.357,0.487,1.0}{$\bullet$} \\ 
O6V & 37500 & 4.0 & 0.359,0.487,1.0 & \#5b7cff & \LARGE \textcolor[rgb]{0.359,0.487,1.0}{$\bullet$} \\ 
O5V & 40000 & 4.0 & 0.358,0.486,1.0 & \#5b7bff & \LARGE \textcolor[rgb]{0.358,0.486,1.0}{$\bullet$} \\ 
O4V & 42500 & 4.0 & 0.357,0.485,1.0 & \#5a7bff & \LARGE \textcolor[rgb]{0.357,0.485,1.0}{$\bullet$} \\ 
O4V & 45000 & 4.0 & 0.357,0.485,1.0 & \#5a7bff & \LARGE \textcolor[rgb]{0.357,0.485,1.0}{$\bullet$} \\ 
O3V & 47500 & 4.0 & 0.357,0.485,1.0 & \#5b7bff & \LARGE \textcolor[rgb]{0.357,0.485,1.0}{$\bullet$} \\ 
O2V & 50000 & 4.0 & 0.358,0.486,1.0 & \#5b7bff & \LARGE \textcolor[rgb]{0.358,0.486,1.0}{$\bullet$} \\ 
O2V & 52500 & 4.0 & 0.359,0.487,1.0 & \#5b7cff & \LARGE \textcolor[rgb]{0.359,0.487,1.0}{$\bullet$} \\ 
O1V & 55000 & 4.0 & 0.361,0.489,1.0 & \#5c7cff & \LARGE \textcolor[rgb]{0.361,0.489,1.0}{$\bullet$} \\ 

\hline
\end{tabular}
\end{table*}

\nocite{*}
\bibliography{ms.bib}%

%

\end{document}